\documentclass[a4paper,12pt]{article}

\usepackage{amsmath}
\usepackage{amssymb}
\usepackage{slashed}

\numberwithin{equation}{section}
\def\appendix#1{\addtocounter{section}{1}\setcounter{equation}{0}
\renewcommand{\thesection}{\Alph{section}}
\section*{Appendix \thesection\protect\indent \parbox[t]{11.15cm}{#1}}
\addcontentsline{toc}{section}{Appendix \thesection\ \ \ #1}}

\begin{document}

\begin{titlepage}
\begin{center}

\vspace*{-1.0cm}

\hfill  DMUS-MP-23-08
\\
\vspace{2.0cm}

\renewcommand{\thefootnote}{\fnsymbol{footnote}}
{\Large{\bf Real Supersymmetric Solutions of $(3,2)$ Signature Five-Dimensional
Supergravity}}
\vskip1cm
\vskip 1.3cm
D. Farotti$^1$, J. B. Gutowski$^1$  and W. A. Sabra$^2$
\vskip 1cm
{\small{\it
$^1$Department of Mathematics, School of Mathematics and Physics,
\\
University of Surrey,
Guildford, GU2 7XH, UK.\\}}
\vskip .6cm {\small{\it
$^2$ Physics Department \\ 
American University of Beirut\\ Lebanon  \\}}
\end{center}
\bigskip
\begin{center}
{\bf Abstract}
\end{center}

We classify supersymmetric solutions of $D=5$ $(3,2)$ signature supergravity with 
either vanishing or imaginary gauge coupling
constant preserving
the minimal $N=2$ supersymmetry. We prove that
the geometry of such solutions is characterized by a nilpotent integrable
endomorphism, and obtain the necessary and sufficient conditions on the
fluxes imposed by supersymmetry. We also construct examples of supersymmetric
domain wall solutions for which the nilpotent integrable endomorphism
is associated with a Killing-Yano 2-form, as well as a new
descendant ``preon" solution which preserves $N=6$ supersymmetry.
This is notable as such $N=6$ descendant solutions do not exist in the standard signature $D=5$ supergravity.

\end{titlepage}

\section{Introduction}

    The classification and analysis of supersymmetric solutions of five-dimensional $N = 2$ supergravity theories coupled to vector multiplets \cite{GST1, GST2} have attracted considerable attention in recent years. A large class of these theories can be obtained via the dimensional reduction of eleven-dimensional supergravity of \cite{CJS} on a Calabi-Yau threefold (CY3) \cite{CDS}. The first systematic classification of supersymmetric solutions of  five-dimensional ungauged minimal supergravity in Lorentzian signature (and without vector multiplets) was given in \cite{Gauntlett:2002nw}. The generalizations of the results of \cite{Gauntlett:2002nw} to solutions of supergravity theories with vector multiplets was later performed in \cite {Jan2004}. Spinorial geometry techniques were then developed in
the context of the classification of supersymmetric supergravity solutions in  ten and eleven dimensions \cite{spinorial}. Such methods have been applied in the classification of many different types of solutions in numerous theories (see \cite{report} and references therein). 
     
     In recent years, there has been renewed interest in the construction and study of supergravity theories in various space-time signatures. Theories with exotic space-time signatures arise naturally in the study of the web of duality symmetries in string and M-theory. Such exotic theories are obtained through a time-like T-duality of ordinary string theories and have unconventional effective field theory description which involves RR fields with wrong sign of kinetic energy terms. Therefore the existence of such theories as a component of the full M-theory is linked to allowing T-duality transformations along a time-like circle \cite{Hull1, Hull2, DHJV}. Non-Lorentzian signatures are also relevant to the study of non-perturbative effects as well as the study of quantum gravity. 
     
     Euclidean versions of the ungauged supergravity theories of \cite{GST1} were considered in \cite{SV}. It was concluded that the kinetic terms of the gauge fields in the Euclidean theory have the non-conventional sign in the Lagrangian of the theory. Five-dimensional theories with various space-time signature can be obtained through the reduction of Hull's exotic eleven-dimensional supergravity on CY3. We shall denote the space-time signature by $(s, t)$ where $s$ refers to spatial dimensions and $t$ refers to time dimensions. The reduction of supergravity theories with signatures $(10, 1)$, $(6, 5)$ and $(2, 9)$ produces five-dimensional theories with signatures $(4, 1)$, $(0, 5)$ and $(2, 3)$ with standard sign of gauge kinetic terms. The reduction of theories with signatures $(1, 10)$, $(5, 6)$ and $(9, 2)$ produces five-dimensional theories with signatures $(1, 4)$, $(5, 0)$ and $(3, 2)$ with the non-conventional sign of the gauge kinetic terms \cite{reduce}. For a detailed analysis of off-shell supersymmetry transformations and associated supersymmetric Lagrangians for five-dimensional vector multiplets in arbitrary space-time dimensions, we refer the reader to \cite{GM}. Exotic gauged supergravity theories can be obtained via non-linear Kaluza Klein reduction of exotic type IIB and M theory \cite{JWW}.

  Solutions for a particular choice of a Dirac Killing spinor were considered for all signatures in ungauged five-dimensional supergravity theories in \cite{para}. It was found that the four-dimensional base space of the Euclidean theories are given, as in the Lorentzian cases, in terms of hyper-K\"ahler manifolds. In terms of supersymmetry, the ungauged $(3,2)$ signature theory admits Majorana Killing spinors, though the spinor $\epsilon$ considered
in \cite{para} was not Majorana. Consequently, such solutions actually preserved an extended $N=4$ supersymmetry, and could be described in terms of a fibration over a 4-dimensional base space
which is equipped with a hypersymplectic structure \cite{Hitchin}. Supersymmetric solutions with Majorana Killing spinors in four-dimensional minimal ungauged $(2,2)$ signature supergravity were classified in \cite{Gutowski:2019hnl}.

The goal of our present work is to classify minimally supersymmetric ($N=2$) solutions of $D=5$, $(3,2)$ supergravity with either vanishing or imaginary gauge coupling constant, for which the Killing spinor is Majorana. In such a case, we shall determine the geometric structures common to all such solutions, and which also underpin the hypersymplectic structures found for the case of enhanced $N=4$ supersymmetry considered in \cite{para}. We shall show that the geometric structure associated with minimal $N=2$ supersymmetry in these theories is characterized by a nilpotent integrable endomorphism, and we fully determine the necessary and sufficient conditions on the fluxes for supersymmetry. We remark that this situation, with respect to the amount of supersymmetry preserved, is novel to theories with such mixed signature, and does not hold for the ungauged $(4,1)$ signature theory considered in \cite{Gauntlett:2002nw}, where the solutions preserve either $N=0$, $N=4$, or $N=8$ supersymmetry, and there are no $N=2$ supersymmetric solutions. 

We also construct a number of examples of supersymmetric solutions, including domain wall solutions, and a novel class of ``descendant" supersymmetric
solutions in the non-minimal ungauged $(3,2)$ theory for which the gravitino Killing spinor equation preserves $N=8$ maximal supersymmetry, but the $N=8$ supersymmetry is broken down to $N=6$ by the gaugino Killing spinor equation. Descendant solutions were first classified in heterotic supergravity \cite{Gran:2007fu, Gran:2007kh}. In five dimensions, the existence of such solutions preserving $N=6$ supersymmetry is again particular to the $(3,2)$  signature; it is absent in the case of $(4,1)$ signature.

The plan of this paper is as follows. In Section 2 we summarize some key properties of the $(3,2)$ signature theories whose supersymmetric solutions we classify. We also use spinorial geometry techniques to determine a simple canonical form for a single Majorana Killing spinor, the corresponding gauge-invariant spinor bilinears, and the associated stabilizer subgroup.
In Section 3 we obtain the necessary and sufficient conditions for $N=2$ supersymmetry obtained from the gravitino and gaugino Killing spinor equations, and prove that the geometric structure underpinning all such solutions is characterized by a nilpotent integrable endomorphism.
In Section 4, we construct some examples, considering solutions for which the spinor bilinear is parallel, and also domain wall solutions for which all the gauge field terms vanish. We also construct the novel $N=6$ descendant
solution mentioned above, which is also a special case of a solution with parallel spinor bilinear. In Section 5 we present our conclusions, as well as a brief discussion 
on integrability conditions of the Killing spinor equations, and enhanced supersymmetry. 
There are four Appendices. In Appendix A we present our Clifford algebra conventions. In Appendix B we list the linear system associated with the gravitino Killing spinor equation, which is used to establish that the necessary conditions which are obtained from the covariant derivative 
of the spinor bilinear are in fact also sufficient for $N=2$  supersymmetry in the gravitino equation. In Appendix C we present some further steps in the analysis of the necessary and sufficient conditions for $N=2$ supersymmetry in the gaugino equation, and which are also useful in considering the integrability conditions of the Killing spinor equations. In Appendix D we present the analysis which is used to construct simple canonical forms for solutions with enhanced $N=4$ supersymmetry, which relates to the classification constructed for a class of $N=4$ supersymemtric solutions in \cite{para}.

\section{Majorana Killing Spinors}

In this section we examine the Killing Spinor Equations (KSE) of five-dimensional
gauged supergravity in signature $(+,+,+,-,-)$ coupled to arbitrary many vector multiplets.
The gravitino equation is 

\begin{eqnarray}
\label{graveq}
{\cal{D}}_\mu \epsilon \equiv \nabla_\mu \epsilon + {k \over 8} \gamma_\mu{}^{\lambda_1 \lambda_2}  
H_{\lambda_1 \lambda_2} \epsilon -{k \over 2} H_{\mu \lambda} \gamma^\lambda \epsilon
-{i \over 2} g X \gamma_\mu \epsilon +{3i \over 2} kg A_\mu \epsilon =0
\end{eqnarray}
where $k$, $g$ are constants and $\nabla$ denotes the Levi-Civita connection. The gaugino KSE
are 
\begin{eqnarray}
\label{gaugino}
{\cal{A}}^I \epsilon \equiv \bigg( k ({\slashed{F}}^I - X^I {\slashed{H}}) -2 {\slashed \partial} X^I + 4ig V_J(X^I X^J - {3 \over 2}Q^{IJ}) \bigg) \epsilon =0 \ .
\end{eqnarray}
The 2-form $H$ is related to the 2-form field strengths $F^I$ by $H= X_I F^I$, with $F^I = d A^I$, $V_J$ are constants, with 
$A= V_I A^I$ and $X= V_I X^I$. The scalar fields $X^I$ are real, and satisfy the condition
\begin{eqnarray}
X_I X^I=1
\end{eqnarray}
where $X_I = {1 \over 6} C_{IJK} X^J X^K$ and the (real) intersection numbers $C_{IJK}$ are
symmetric in $I, J, K$. Further details of this formalism can be found in \cite{Gutowski:2007ai}.

We shall proceed to consider the constants $k$ and $g$ appearing in these KSE. We first remark that the constants appearing in ({\ref{graveq}}) have already been fixed in such a way as to ensure that the integrability conditions obtained by computing $\gamma^\mu [{\mathcal{D}}_\mu,{\mathcal{D}}_\nu] \epsilon$ can be expanded out solely in terms of field equations and Bianchi identities. In particular, modifying the coefficient of the $A_\mu$ term in ({\ref{graveq}}) would produce from such a calculation a term linear in the gauge field strength which would not correspond to such a field equation or Bianchi identity. Furthermore, it is also straightforward to show that one must also take $k \in \mathbb{R}$. To see this, it is necessary to consider the integrability conditions in more detail. These have already been computed for gauged supergravity coupled to vector multiplets in \cite{Gutowski:2007ai} for the case of the theory with signature $(+,-,-,-,-)$; moreover the gamma matrices in that theory
were also taken to satisfy the same conditions given in ({\ref{orient}}). It follows that the integrability conditions for the theory considered in this paper can be directly read off from those
given in \cite{Gutowski:2007ai}, on making the replacements $F^I \rightarrow k F^I$, $A^I \rightarrow k A^I$, $A \rightarrow kA$, and $\chi \rightarrow -g$. In particular, the
integrability conditions, assuming the Bianchi identities $dF^I=0$, imply that
\begin{eqnarray}
\label{scalint}
\bigg( S_I -{2 \over 3} (G_{I \mu} - X_I X^J G_{J \mu} ) \gamma^\mu \bigg) \epsilon =0
\end{eqnarray}
and
\begin{eqnarray}
\label{einstint}
\bigg(E_{\mu \nu} \gamma^\nu +{1 \over 3} X^I (\gamma_\mu{}^\nu G_{I \nu}
- 2G_{I \mu}) \bigg) \epsilon =0
\end{eqnarray}
where $E_{\mu \nu}$ and $S_I$ correspond to Einstein and scalar field equation terms
respectively, and
\begin{eqnarray}
G_{I \mu} = k \nabla^\nu \big(Q_{IJ} F^J_{\mu \nu}\big) -{1 \over 16} k^2
C_{IJK} \epsilon_\mu{}^{\nu_1 \nu_2 \nu_3 \nu_4} F^J_{\nu_1 \nu_2} F^K_{\nu_3 \nu_4} \ .
\end{eqnarray}
As we require that $G_{I \mu}=0$ should correspond to a gauge field equation obtained
from a real Einstein-Maxwell-Chern-Simons action, we therefore impose that $k \in \mathbb{R}$,
and without loss of generality we therefore set $k=1$.

It then remains to consider $g$. If $g \in \mathbb{R}$ with $g \neq 0$, then requiring that ({\ref{gaugino}}) admits a Majorana Killing spinor $\epsilon$ implies that
\begin{eqnarray}
X X_I = V_I \ .
\end{eqnarray}
In turn, this implies that $\partial_\mu X X_I + X \partial_\mu X_I =0$, and on contracting with
$X^I$ this implies that $X$ is constant, so in this case $dH=0$. Furthermore, the gravitino equation also factorizes and implies that
\begin{eqnarray}
\nabla_\mu \epsilon + {1 \over 8} \gamma_\mu{}^{\lambda_1 \lambda_2}  
H_{\lambda_1 \lambda_2} \epsilon -{1 \over 2} H_{\mu \lambda} \gamma^\lambda \epsilon=0
\end{eqnarray}
and hence it follows that the geometric conditions, and the conditions on $H$, obtained
from the gravitino equation correspond to a special case of those obtained by considering
those obtained from the analysis of the minimal ungauged supergravity (with additional conditions
on the flux).

Conversely, if $ig \in \mathbb{R}$ (including the case $g=0$),
then $[C*, {\mathcal{D}}_\mu]=0$ and $[C*, {\mathcal{A}}^I]=0$. In particular, for such choices of $g$, if $\epsilon$ is a Killing spinor then so is $C* \epsilon$. Hence it follows that for solutions preserving the minimal $N=2$ supersymmetry, the Killing spinors must be of the form
$\{ \epsilon, i \epsilon \}$ where $\epsilon$ is a Majorana spinor satisfying $C* \epsilon = \epsilon$. We shall in this work concentrate on the case for which $ig \in \mathbb{R}$ (including the case $g=0$), and for which solutions with minimal supersymmetry are described by a single Majorana Killing spinor. 

\subsection{Majorana Spinor Orbits, Bilinears, and Stabilizer}

To proceed, we next consider how to apply $Spin(3,2)$ gauge transformations in order
to simplify a (single) Majorana spinor. A general spinor $\epsilon$ satisfying 
$\epsilon = C* \epsilon$ is given by
\begin{eqnarray}
\label{maj1}
\epsilon = \lambda .1 + {\bar{\lambda}} e_{12} + \mu e_1 + {\bar{\mu}} e_2
\end{eqnarray}
for $\lambda \in {\mathbb{C}}$, $\mu \in {\mathbb{C}}$.
The $Spin(3,2)$ gauge transformations generated by
$\{ \gamma_{12}-\gamma_{34}, \gamma_{13}-\gamma_{24}, \gamma_{14}+\gamma_{23} \}$
leave $\{e_1, e_2\}$ invariant and act as $\{\sigma_1, \sigma_2, i \sigma_3\}$
on $\{1, e_{12} \}$. Also, the $Spin(3,2)$ gauge transformations generated by
$\{ \gamma_{12}+\gamma_{34}, \gamma_{13}+\gamma_{24}, \gamma_{14}-\gamma_{23} \}$
leave $\{1, e_{12}\}$ invariant and act as $\{\sigma_1, \sigma_2, i \sigma_3\}$
on $\{e_1, e_2 \}$. It follows that by choosing gauge transformations
generated by $\gamma_{14}+\gamma_{23}$ and $\gamma_{14}-\gamma_{23}$
we can without loss of generality set $\lambda \in {\mathbb{R}}, \mu \in {\mathbb{R}}$ in ({\ref{maj1}}). Furthermore, by then utilizing gauge transformations
generated by $\gamma_{12}-\gamma_{34}$ and $\gamma_{12}+\gamma_{34}$ we can without
loss of generality take $\lambda, \mu \in \{-1, 0, 1\}$. It therefore appears that there are three possible canonical forms for such a Majorana spinor

\begin{itemize}
\item[(i)] $\epsilon = 1+e_{12}$
\item[(ii)] $\epsilon = e_1 + e_2$
\item[(iii)] $\epsilon = 1+e_{12} \pm (e_1 + e_2)$
\end{itemize}

However as $e^{{\pi \over 2} \gamma_{51}} (e_1 + e_2) = 1+e_{12}$, it follows that
$e_1 + e_2$ is in the same orbit as $1+e_{12}$. Furthermore, we also have
\begin{eqnarray}
e^{\pm {1 \over 2} \gamma_5(\gamma_1 + \gamma_4)} (1+e_{12} \pm (e_1 + e_2))
= 1+e_{12}
\end{eqnarray}
and so it follows that $1+e_{12} \pm (e_1 + e_2)$ is also in the same orbit as
$1+e_{12}$. So, there is a single Majorana spinor orbit. Hence, without loss of generality, by using appropriately
chosen $Spin(3,2)$ gauge transformations, we can take 
a single Majorana spinor to be written as
\begin{eqnarray}
\label{maj2}
\epsilon = 1+e_{12} \ .
\end{eqnarray}

We remark that there are only two $Spin(3,2)$ gauge-invariant spinor bilinears associated with this spinor. There is a 2-form $\omega$ with components
\begin{eqnarray}
\label{omdef}
\omega_{\mu \nu} = {\mathcal{B}} (\epsilon, \gamma_{\mu \nu} \epsilon)
\end{eqnarray}
where the ${\mathcal{B}}$ is defined by ({\ref{ginp}}). One finds that
\begin{eqnarray}
\label{omdef2}
\omega= 2 \tau \wedge {\bar{\tau}}, \qquad {\rm where} \qquad  \tau = {\bf{e}}^1 + i {\bf{e}}^{\bar{2}} \ .
\end{eqnarray}
The other spinor bilinear is the 3-form corresponding to the Hodge dual of $\omega$,
\begin{eqnarray}
\star \omega = - {\bf{e}}^5 \wedge \omega \ .
\end{eqnarray}
There are no other non-zero $Spin(3,2)$ gauge-invariant spinor bilinears 
associated with $\epsilon$.

It is also useful to consider the stability subgroup of $Spin(3,2)$ which leaves invariant this spinor. We shall solve the condition
\begin{eqnarray}
\lambda^{\mu \nu} \gamma_{\mu \nu} \epsilon =0 \ .
\end{eqnarray}
The following conditions on $\lambda^{\mu \nu}$ are obtained:
\begin{eqnarray}
\lambda^{51} + \lambda^{54}=0, \quad \lambda^{52} + \lambda^{53}=0,
\quad \lambda^{12} - \lambda^{34} + \lambda^{13} - \lambda^{24}=0, \quad
\lambda^{14}+ \lambda^{23}=0 \ .
\nonumber \\
\end{eqnarray}

Hence, the stabilizer subgroup is 6-dimensional, and it will be convenient to
use the following basis
\begin{eqnarray}
K_1 = \gamma_{14}-\gamma_{23}, \quad K_2 = \gamma_{12}+\gamma_{34}+\gamma_{24}+\gamma_{13},
\quad K_3 = \gamma_{12}+\gamma_{34}-\gamma_{24}-\gamma_{13}
\nonumber \\
\end{eqnarray}
and
\begin{eqnarray}
S_1 = \gamma_{51}-\gamma_{54}, \quad S_2 = \gamma_{52} - \gamma_{53}, \quad
S_3 = -\gamma_{12}+\gamma_{13} - \gamma_{24} + \gamma_{34} \ .
\end{eqnarray}

It is straightforward to show that the $S_i$ satisfy the Heisenberg algebra, ${\mathfrak{n}}_3$,
with the only non-zero $[S, S]$ commutator being given by
\begin{eqnarray}
[S_1, S_2]=2S_3
\end{eqnarray}
and the $K_j$ satisfy the ${\mathfrak{sl}}(2, \mathbb{R})$ algebra, with the non-zero $[K, K]$
commutators given by
\begin{eqnarray}
[K_1, K_2]=4 K_2, \quad [K_3, K_1]=4 K_3, \quad [K_2, K_3]=-8 K_1 \ .
\end{eqnarray}
The remaining non-zero $[S,K]$ commutators are given by
\begin{eqnarray}
[S_1, K_1]=-2S_1, \quad [S_1, K_3]=4 S_2, \quad [S_2, K_1]=2S_2, \quad [S_2, K_2]=-4 S_1 \ .
\end{eqnarray}
Hence, the stabilizer subgroup is the semidirect sum ${\mathfrak{sl}}(2, \mathbb{R})
\ltimes {\mathfrak{n}}_3$.

\section{Analysis of the KSE}

In this section we analyse the necessary and sufficient conditions for supersymmetry
obtained from the KSE ({\ref{graveq}}) and ({\ref{gaugino}}), for the case $k=1$, $ig \in \mathbb{R}$,
taking the spinor $\epsilon$ to be Majorana, with $\epsilon=1+e_{12}$. As noted in the previous
section, there is a single nontrivial spinor bilinear $\omega$ ({\ref{omdef}}) (as well as the 
Hodge dual of $\omega$).

\subsection{Gravitino KSE}

We begin with the gravitino KSE ({\ref{graveq}}). On taking the covariant derivative of $\omega$, ({\ref{graveq}}) implies the following condition
\begin{eqnarray}
\label{eqx4}
\nabla_\alpha \omega_{\mu \nu} &=& (\star \omega)_\alpha{}^\lambda{}_{[\nu} H_{\mu] \lambda} -{1 \over 2} H^{\lambda_1 \lambda_2} (\star \omega)_{\lambda_1 \lambda_2 [\nu}
\eta_{\mu] \alpha} - H_{\alpha \lambda} (\star \omega)^\lambda{}_{\mu \nu}
\nonumber \\
&-& i g X (\star \omega)_{\alpha \mu \nu}
- 3i  g A_\alpha \omega_{\mu \nu} \ .
\end{eqnarray}
This condition is a necessary condition obtained from ({\ref{graveq}}); in fact it can also be shown that it is equivalent to ({\ref{graveq}}). This follows from decomposing ({\ref{eqx4}}) into components, and comparing the resulting equations to the linear system listed in ({\ref{glinsys}}), which is equivalent to ({\ref{graveq}}). ({\ref{eqx4}}) and ({\ref{glinsys}}) are equivalent, which implies that ({\ref{eqx4}}) is a necessary and sufficient condition for the gravitino KSE to hold.

We proceed then to analyse the content of ({\ref{eqx4}}). On setting $\mu=p$, $\nu=q$, $\alpha=r$ in ({\ref{eqx4}}) one obtains

\begin{eqnarray}
\label{ex1}
2 (\nabla_r \tau_p) {\bar{\tau}}_q + 2 (\nabla_r {\bar{\tau}}_q) \tau_p
-2 (\nabla_r \tau_q) {\bar{\tau}}_p - 2 (\nabla_r {\bar{\tau}}_p) \tau_q
&=&  (\tau_r {\bar{\tau}}_q - \tau_q {\bar{\tau}}_r) H_{p5}
\nonumber \\
&-&  (\tau_r {\bar{\tau}}_p - \tau_p {\bar{\tau}}_r) H_{q5}
\nonumber \\
&-& (\tau^\ell {\bar{\tau}}_q - \tau_q {\bar{\tau}}^\ell)
H_{\ell 5} \eta_{pr}
\nonumber \\
&+& (\tau^\ell {\bar{\tau}}_p - \tau_p {\bar{\tau}}^\ell)
H_{\ell 5} \eta_{qr}
\nonumber \\
&+&2 H_{r5} (\tau_p {\bar{\tau}}_q - \tau_q {\bar{\tau}}_p)
\nonumber \\
&-&6i g A_r  (\tau_p {\bar{\tau}}_q - \tau_q {\bar{\tau}}_p) \ .
\nonumber \\
\end{eqnarray}
On setting
\begin{eqnarray}
B_{rp} &=& 2 \nabla_r \tau_p - \tau_r H_{p5}
-\tau_p H_{r5} + \tau^\ell H_{\ell 5} \eta_{pr} + 3i g A_r \tau_p
\end{eqnarray}
the condition ({\ref{ex1}}) can be rewritten as
\begin{eqnarray}
B_{rp} {\bar{\tau}}_q +{\bar{B}}_{rq} \tau_p - B_{rq} {\bar{\tau}}_p
- {\bar{B}}_{rp} \tau_q =0 \ .
\end{eqnarray}
It follows that ({\ref{ex1}}) is equivalent to
\begin{eqnarray}
 2 \nabla_r \tau_p - \tau_r H_{p5}
-\tau_p H_{r5} +
\tau^\ell H_{\ell 5} \eta_{pr} + 3i g A_r \tau_p
= U_r \tau_p + W_r {\bar{\tau}}_p
\end{eqnarray}
where $U$ is imaginary, and $W$ is complex. 

On setting $\mu=5$, $\nu=p$, $\alpha=5$ in ({\ref{eqx4}}) one obtains
\begin{eqnarray}
2 \nabla_5 \tau_5 + 2 \tau^\ell H_{5 \ell} =0 \ .
\end{eqnarray}

 On setting $\mu=p$, $\nu=q$, $\alpha=5$ in ({\ref{eqx4}}) one obtains
\begin{eqnarray}
2 (\nabla_5 \tau_p) {\bar{\tau}}_q + 2 (\nabla_5 {\bar{\tau}}_q) \tau_p
-2 (\nabla_5 \tau_q) {\bar{\tau}}_p -2 (\nabla_5 {\bar{\tau}}_q) \tau_q
&=& -(\tau^\ell {\bar{\tau}}_q - \tau_q {\bar{\tau}}^\ell)
H_{p \ell}
\nonumber \\
&+&(\tau^\ell {\bar{\tau}}_p - \tau_p {\bar{\tau}}^\ell)
H_{q \ell}
\nonumber \\
&+&(2ig X -6ig A_5)
(\tau_p {\bar{\tau}}_q - \tau_q {\bar{\tau}}_p) \ .
\nonumber \\
\end{eqnarray}
On setting 
\begin{eqnarray}
Z_p = 2 \nabla_5 \tau_p +  \tau^\ell H_{p \ell}
+(-igX +3ig A_5) \tau_p
\end{eqnarray}
one obtains
\begin{eqnarray}
Z_p {\bar{\tau}}_q + {\bar{Z}}_q \tau_p - Z_q {\bar{\tau}}_p - {\bar{Z}}_p \tau_q =0
\end{eqnarray}
which implies that
\begin{eqnarray}
2 \nabla_5 \tau_p + \tau^\ell H_{p \ell}
+(-igX +3ig A_5) \tau_p
= \alpha \tau_p + \beta {\bar{\tau}}_p
\end{eqnarray}
where $\alpha$ is imaginary, and $\beta$ is complex. The remaining case corresponds
to setting $\mu=5$, $\nu=p$, $\alpha=q$ in ({\ref{eqx4}}) which produces the
condition

\begin{eqnarray}
\label{auxcond1a}
2 (\nabla_q \tau_5) {\bar{\tau}}_p - 2 (\nabla_q {\bar{\tau}}_5) \tau_p
&=&  (\tau_q {\bar{\tau}}^\ell - \tau^\ell {\bar{\tau}}_q)
H_{p \ell} - \tau^{\ell_1} {\bar{\tau}}^{\ell_2} H_{\ell_1 \ell_2} \eta_{pq} 
\nonumber \\
&+&2(\tau_p {\bar{\tau}}^\ell - \tau^\ell {\bar{\tau}}_p) H_{q \ell} +2ig X (\tau_p {\bar{\tau}}_q - \tau_q {\bar{\tau}}_p) \ .
\nonumber \\
\end{eqnarray}

To proceed, antisymmetrize ({\ref{auxcond1a}}) in $p,q$. On defining
\begin{eqnarray}
C_p = 2 \nabla_p \tau_5 +\tau^\ell H_{p \ell}
+2igX \tau_p
\end{eqnarray}
the resulting condition is
\begin{eqnarray}
C_q {\bar{\tau}}_p - {\bar{C}}_q \tau_p - C_p {\bar{\tau}}_q + {\bar{C}}_p \tau_q =0
\end{eqnarray}
which implies that
\begin{eqnarray}
2 \nabla_p \tau_5 +\tau^\ell H_{p \ell}
+2igX \tau_p = \Lambda \tau_p + \Theta {\bar{\tau}}_p
\end{eqnarray}
where $\Lambda+ {\bar{\Lambda}}=0$. Next, substitute this condition back into
({\ref{auxcond1a}}) to find

\begin{eqnarray}
\label{auxcond1b}
\Lambda (\tau_q {\bar{\tau}}_p + \tau_p {\bar{\tau}}_q)
+ \Theta {\bar{\tau}}_p {\bar{\tau}}_q - {\bar{\Theta}} \tau_p \tau_q
&=& (\tau_q {\bar{\tau}}^\ell - \tau^\ell {\bar{\tau}}_q)
H_{p \ell}
\nonumber \\
&+& (\tau_p {\bar{\tau}}^\ell - \tau^\ell {\bar{\tau}}_p)
H_{q \ell} 
\nonumber \\
&-& \tau^{\ell_1} {\bar{\tau}}^{\ell_2} H_{\ell_1 \ell_2} \eta_{pq} \ .
\end{eqnarray}

It will be convenient to write
\begin{eqnarray}
\label{auxcond1c}
2 \tau^\ell H_{p \ell} = \beta_1 \tau_p + \beta_2 {\bar{\tau}}_p
+ \beta_3 \xi_p
\end{eqnarray}
where here $\xi = {\bf{e}}^1 -i {\bf{e}}^{\bar{2}}$, satisfies $\xi. \xi = {\bar{\xi}}.{\bar{\xi}}= \xi. \tau=0$ and $\xi . {\bar{\tau}} = 2$ so that the metric
satisfies
\begin{eqnarray}
\label{auxcond1d}
\eta_{pq} = {1 \over 2} \tau_q {\bar{\xi}}_p +{1 \over 2} \tau_p {\bar{\xi}}_q
+ {1 \over 2} {\bar{\tau}}_q \xi_p + {1 \over 2} {\bar{\tau}}_p \xi_q \ .
\end{eqnarray} 
On substituting ({\ref{auxcond1c}}) into ({\ref{auxcond1b}}), one finds the conditions
\begin{eqnarray}
2 \Lambda = {\bar{\beta}}_1 - \beta_1 \ , \qquad \Theta= - \beta_2 \ , \qquad
\beta_3 + {\bar{\beta}}_3 =0 \ .
\end{eqnarray}

On collating all of the conditions, ({\ref{eqx4}}) is equivalent to:

\begin{eqnarray}
\label{block3a}
2 \nabla_r \tau_p -\tau_r H_{p5} - \tau_p H_{r5} + \tau^m H_{m5} \eta_{pr}+3igA_r \tau_p = U_r \tau_p + W_r {\bar{\tau}}_p
\end{eqnarray}
\begin{eqnarray}
\label{block3b}
\nabla_5 \tau_5 = (i_\tau H)_5 \ , \qquad \nabla_5 \tau_p = {1 \over 2}(i_\tau H)_p
+{\alpha \over 2} \tau_p + {\beta \over 2}{\bar{\tau}}_p
\end{eqnarray}
\begin{eqnarray}
\label{block3c}
2 \nabla_p \tau_5 + \big(\beta_1 - {1 \over 2} {\bar{\beta}}_1 + 2igX) \tau_p
+{3 \over 2} \beta_2 {\bar{\tau}}_p + {1 \over 2} \beta_3 \xi_p =0
\end{eqnarray}
\begin{eqnarray}
\label{block3d}
2 \tau^\ell H_{p \ell} = \beta_1 \tau_p + \beta_2 {\bar{\tau}}_p
+ \beta_3 \xi_p
\end{eqnarray}
where $U_r$, $\beta_3$ are imaginary; $W_r$, $\beta_1$ and $\beta_2$ are complex.

There remains some gauge freedom in this system. In particular, one may define
${\hat{\tau}} = {\bar{\mu}} \tau - \nu {\bar{\tau}}$ for $\mu, \nu \in \mathbb{C}$ such that $|\mu|^2 - |\nu|^2=1$. With this choice
\begin{eqnarray}
\tau \wedge {\bar{\tau}} = {\hat{\tau}} \wedge {\bar{\hat{\tau}}} \ .
\end{eqnarray}
Moreover, we set ${\hat{\xi}} = {\bar{\mu}} \xi + \nu {\bar{\xi}}$, which ensures that
the condition ({\ref{auxcond1d}}) holds, with $\tau, \xi$ replaced with ${\hat{\tau}}, {\hat{\xi}}$. It remains to consider the effect of such a frame redefinition on the conditions ({\ref{block3a}})-({\ref{block3d}}).
The form of the first condition in ({\ref{block3b}}) is invariant under such a redefinition.
 After some calculation, it follows that the form of ({\ref{block3a}}) is preserved, with $U_r, W_r$ replaced with ${\hat{U}}_r$, ${\hat{W}}_r$ where
\begin{eqnarray}
{\hat{U}}_r &=& 2 (\mu \nabla_r {\bar{\mu}} - {\bar{\nu}} \nabla_r \nu)
+ |\mu|^2 U_r - |\nu|^2 {\bar{U}}_r + {\bar{\mu}} {\bar{\nu}} W_r
- \mu \nu {\bar{W}}_r
\nonumber \\
{\hat{W}}_r &=& 2(\nu \nabla_r {\bar{\mu}} - {\bar{\mu}} \nabla_r \nu)
+ {\bar{\mu}} \nu (U_r - {\bar{U}}_r) + {\bar{\mu}}^2 W_r - \nu^2 {\bar{W}}_r \ .
\end{eqnarray}
Next we consider ({\ref{block3d}}); the form of this condition is unchanged, with
$\tau, \xi$ replaced with ${\hat{\tau}}, {\hat{\xi}}$, and $\beta_1$, $\beta_2$, 
$\beta_3$ replaced with

\begin{eqnarray}
{\hat{\beta}}_1 &=& |\mu|^2 \beta_1 - |\nu|^2 {\bar{\beta}}_1 - \mu \nu {\bar{\beta}}_2
+ {\bar{\mu}} {\bar{\nu}} \beta_2
\nonumber \\
{\hat{\beta}}_2 &=& \bar{\mu}^2 \beta_2 - \nu^2 {\bar{\beta}}_2 
+ {\bar{\mu}} \nu (\beta_1 - {\bar{\beta}}_1)
\nonumber \\
{\hat{\beta}}_3 &=& \beta_3
\end{eqnarray}
and furthermore, the form of ({\ref{block3c}}) is also unchanged under the frame redefinition, again with $\tau, \xi$ replaced with ${\hat{\tau}}, {\hat{\xi}}$, and $\beta_1$, $\beta_2$, 
$\beta_3$ replaced with ${\hat{\beta}}_1$, ${\hat{\beta}}_2$, ${\hat{\beta}}_3$.
It remains to consider the second condition in ({\ref{block3b}}), which is preserved with $\tau, \xi$ replaced with ${\hat{\tau}}, {\hat{\xi}}$; $\beta_1$, $\beta_2$, 
$\beta_3$ replaced with ${\hat{\beta}}_1$, ${\hat{\beta}}_2$, ${\hat{\beta}}_3$;
and $\alpha$, $\beta$ replaced with

\begin{eqnarray}
\label{gaugefreedom1}
{\hat{\alpha}} &=& \alpha (|\mu|^2+|\nu|^2) + \beta {\bar{\mu}} {\bar{\nu}}
- {\bar{\beta}} \mu \nu + 2 \mu \nabla_5 {\bar{\mu}} - 2 {\bar{\nu}} \nabla_5 \nu
\nonumber \\
{\hat{\beta}} &=& \beta {\bar{\mu}}^2 - {\bar{\beta}} \nu^2 +2 {\bar{\mu}} \nu \alpha
+2 \nu \nabla_5 {\bar{\mu}} - 2 {\bar{\mu}} \nabla_5 \nu \ .
\end{eqnarray}

In particular, on using $|\mu|^2 - |\nu|^2=1$, one may choose $\mu$, $\nu$ so that 
${\hat{\alpha}}=0$, ${\hat{\beta}}=0$. Using such a choice of gauge, one may set,
without loss of generality, $\beta$ to be any arbitrary complex value, and 
$\alpha$ to be any arbitrary imaginary value in ({\ref{block3b}}).

To proceed, consider ({\ref{block3a}}). On contracting with ${\bar{\xi}}^p$, one obtains an expression for $U_p$, and also on contracting with 
$\xi^p$ one obtains $W_p$. On imposing that $U_p$ is imaginary, one finds the
following condition 

\begin{eqnarray}
\label{hr5sol}
H_{r5} &=&  {\bar{\xi}}^p \nabla_r \tau_p + \xi^p \nabla_r {\bar{\tau}}_p
+ig\big(6 A_r -2 \tau_r {\bar{\xi}}^\ell A_\ell -2 {\bar{\tau}}_r \xi^\ell A_\ell \big)
\nonumber \\
&-&{1 \over 3} \tau_r \big( \xi^p {\bar{\xi}}^q \nabla_q {\bar{\tau}}_p
+ {\bar{\xi}}^p {\bar{\xi}}^q \nabla_q \tau_p \big)
-{1 \over 3} {\bar{\tau}}_r \big( {\bar{\xi}}^p \xi^q \nabla_q \tau_p
+\xi^p \xi^q \nabla_q {\bar{\tau}}_p \big)
\end{eqnarray}
on substituting this expression back into ({\ref{block3a}}), the remaining
content of this condition is

\begin{eqnarray}
\label{block3a1a}
2 \nabla_r \tau_p &=& \tau_p {\bar{\xi}}^m \nabla_r \tau_m + {\bar{\tau}}_p \xi^m 
\nabla_r \tau_m +3ig \xi_p \big( {\bar{\tau}}^\ell A_\ell \tau_r
-\tau^\ell A_\ell {\bar{\tau}}_r \big)
\nonumber \\
&+&{1 \over 2} \xi_p \tau_r \big( \xi^n {\bar{\tau}}^q \nabla_q {\bar{\tau}}_n
+ {\bar{\xi}}^n {\bar{\tau}}^q \nabla_q \tau_n \big) -{1 \over 2} \xi_p {\bar{\tau}}_r
 \big( {\bar{\xi}}^n \tau^q \nabla_q \tau_n + \xi^n
\tau^q \nabla_q {\bar{\tau}}_n \big) \ .
\nonumber \\
\end{eqnarray}

Furthermore, the condition ({\ref{block3a1a}}) is equivalent to

\begin{eqnarray}
\label{block3a1b} 
6ig {\bar{\tau}}^\ell A_\ell = {\bar{\tau}}^p {\bar{\xi}}^q \nabla_q \tau_p
- \xi^p {\bar{\tau}}^q \nabla_q {\bar{\tau}}_p - {\bar{\xi}}^p {\bar{\tau}}^q \nabla_q \tau_p
\end{eqnarray}
and
\begin{eqnarray}
\label{block3a1c} 
\tau^p \tau^q \nabla_p {\bar{\tau}}_q =0 \ .
\end{eqnarray}

Next, consider the condition ({\ref{block3c}}), which is equivalent to

\begin{eqnarray}
\label{block3c1c}
\tau^p \nabla_p \tau_5 =0
\end{eqnarray}

and

\begin{eqnarray}
\label{extrageocomp1}
\beta_1 &=&  -4igX -{4 \over 3} {\bar{\xi}}^p \nabla_p \tau_5 -{2 \over 3} \xi^p \nabla_p {\bar{\tau}}_5
\nonumber \\
\beta_2 &=& -{2 \over 3} \xi^p \nabla_p \tau_5
\nonumber \\
\beta_3 &=& -2 {\bar{\tau}}^p \nabla_p \tau_5 \ .
\end{eqnarray}

Further conditions are obtained by substituting the expressions
for $\beta_1$, $\beta_2$, $\beta_3$ into ({\ref{block3b}}):

\begin{eqnarray}
\label{extrageocomp2a}
\alpha = {1 \over 2}{\bar{\xi}}^p (\nabla_5 \tau_p-{1 \over 3} \nabla_p \tau_5)
- {1 \over 2} \xi^p (\nabla_5 {\bar{\tau}}_p - {1 \over 3} \nabla_p {\bar{\tau}}_5)
\end{eqnarray}
\begin{eqnarray}
\label{extrageocomp2b}
\beta = \xi^p (\nabla_5 \tau_p -{1 \over 3} \nabla_p \tau_5)
\end{eqnarray}
\begin{eqnarray}
\label{extrageocomp2c}
{\bar{\tau}}^p \nabla_5 \tau_p &=& {\bar{\tau}}^p \nabla_p \tau_5
\end{eqnarray}
and
\begin{eqnarray}
\label{repart}
{\bar{\xi}}^p (d\tau)_{5p} + \xi^p (d \bar{\tau})_{5p} = 4igX \ .
\end{eqnarray}

We remark that ({\ref{extrageocomp2c}})
implies that $\beta_3$ as given in  ({\ref{extrageocomp1}}) is imaginary,
as required. A further geometric condition is obtained from the condition
$\nabla_5 \tau_5 = \tau^\ell H_{\ell 5}$, together with ({\ref{hr5sol}})
and also ({\ref{block3a1b}}):

\begin{eqnarray}
\label{extrageocomp3}
\nabla_5 \tau_5 = -{\bar{\tau}}^p \xi^q \nabla_q \tau_p \ .
\end{eqnarray}

Hence, the geometric conditions (if $g \neq 0$) are ({\ref{block3a1c}}), ({\ref{block3c1c}}),
({\ref{extrageocomp2a}}), ({\ref{extrageocomp2b}}), ({\ref{extrageocomp2c}}), 
({\ref{repart}}) and ({\ref{extrageocomp3}}); the flux conditions (if $g \neq 0$)
are given by ({\ref{hr5sol}}), ({\ref{block3a1b}}) and ({\ref{extrageocomp1}}).

To proceed further, it is useful to utilize a specific gauge choice
associated with ({\ref{gaugefreedom1}}). In particular, one may without loss of generality choose a gauge with respect to which 
\begin{eqnarray}
\alpha = {1 \over 3} {\bar{\xi}}^p \nabla_p \tau_5 - {1 \over 3} \xi^p \nabla_p {\bar{\tau}}_5, \qquad \beta = {2 \over 3} \xi^p \nabla_p \tau_5 \ .
\end{eqnarray}
On utilizing such a gauge, the conditions ({\ref{block3c1c}}), ({\ref{extrageocomp2a}}), ({\ref{extrageocomp2b}}), ({\ref{extrageocomp2c}})
and ({\ref{repart}}) are equivalent to{\footnote{If $V$ is a 1-form, then we denote by $V^\sharp$ the vector field dual to $V$ with respect to the metric.}}

\begin{eqnarray}
\label{lietau}
{\cal{L}}_{({\bf{e}}^5)^\sharp} \tau = igX \tau \ .
\end{eqnarray}

Furthermore, the condition ({\ref{extrageocomp3}}) is equivalent to

\begin{eqnarray}
\label{extrageocomp3b}
\nabla^\mu \tau_\mu = -{1 \over 2} (\xi^\mu {\bar\tau}^\nu + {\bar{\xi}}^\mu
\tau^\nu) (d \tau)_{\mu \nu}
\end{eqnarray}
and the condition ({\ref{block3a1c}}) is equivalent to requiring that the Lee form
$\Theta$ of $\omega$ vanish,
\begin{eqnarray}
\Theta_\lambda \equiv \omega_\lambda{}^\sigma \nabla^\nu \omega_{\nu \sigma} =0 \ .
\end{eqnarray}

Further covariantization of the geometric conditions may be obtained by defining

\begin{eqnarray}
\label{nijen1}
{\widehat{\cal{N}}}^\lambda_{\mu \nu}
&=& - \omega_\sigma{}^\lambda \nabla_\mu \omega_\nu{}^\sigma
+ \omega_\sigma{}^\lambda \nabla_\nu \omega_\mu{}^\sigma
- \omega_\nu{}^\sigma \nabla_\sigma \omega_\mu{}^\lambda
+ \omega_\mu{}^\sigma \nabla_\sigma \omega_\nu{}^\lambda
\nonumber \\
&-& \omega_{\mu \nu} \nabla^\sigma \omega_\sigma{}^\lambda \ .
\end{eqnarray}

It is straightforward to show that all of the geometric conditions  are equivalent to ${\widehat{\cal{N}}}=0$, together with ({\ref{lietau}}). However, we also note that
({\ref{lietau}}) implies that
\begin{eqnarray}
\label{lie2form}
{\cal{L}}_{{({\bf{e}}^5)}^\sharp} \omega = 2igX \omega
\end{eqnarray}
and conversely, if ({\ref{lie2form}}) holds, there exists a gauge in which
({\ref{lietau}}) holds. Hence, the geometric conditions are also equivalent
to ${\widehat{\cal{N}}}=0$ together with ({\ref{lie2form}}).

It is also useful to consider the endomorphism $J: T {\mathcal{M}} \rightarrow T {\mathcal{M}}$
generated by $\omega$ given by
\begin{eqnarray}
g(JX,Y)= -i \omega(X,Y)
\end{eqnarray}
for $X$, $Y \in T {\mathcal{M}}$,
which is nilpotent, $J^2=0$ (though ${\rm Im} J \neq {\rm Ker J}$). It is straightforward to show that the geometric conditions imply that $J$ is integrable, in the sense that if $X, Y \in {\rm Ker} J$ then $[X,Y] \in {\rm Ker} J$. In comparison, for integrability of an almost complex structure $I$, one requires that if $X, Y$ in ${\rm Ker}(1+iI)$ then $[X,Y] \in {\rm Ker} (1+iI)$. Here, the integrability of $J$ differs, presumably due to the fact that $J^2=0$ in contrast to $I^2=-1$. Furthermore, the conditions ${\widehat{\mathcal{N}}}=0$ are equivalent to
\begin{eqnarray}
J[JX,Y] + J [X,JY] -[JX,JY] + \omega(X,Y) (\star d \star \omega)^\sharp =0
\end{eqnarray}
for all vector fields $X$ and $Y$. The Nijenhuis tensor of $J$ \cite{nijenhuis} is defined by
\begin{eqnarray}
N_J(X,Y) &=& J^2[X,Y]+[JX,JY] -J[JX,Y] -J[X,JY] 
\nonumber \\
&=& [JX,JY] -J[JX,Y] -J[X,JY]
\end{eqnarray}
on using $J^2=0$. Hence the conditions ${\widehat{\mathcal{N}}}=0$ are equivalent to
\begin{eqnarray}
\label{njcond}
N_J(X,Y)=\omega(X,Y) (\star d \star \omega)^\sharp \ .
\end{eqnarray}
As has been mentioned, this condition, together with ({\ref{lietau}}) encodes all
of the geometric conditions necessary and sufficient for the supersymmetry.
Notably, the associated Haantjes tensor \cite{haantjes}
\begin{eqnarray}
H_J(X,Y) = J^2 N_J(X,Y) + N_J(JX,JY) -J (N_J(JX,Y)+N_J(X,JY))
\nonumber \\
\end{eqnarray}
vanishes identically, on using $J^2=0$, together with ({\ref{njcond}}). This implies that certain linear combinations
of eigenvectors of $J$ generate integrable distributions
\cite{nijenhuis, haantjes}; it would be interesting to explore
this further.

In addition, we note that ({\ref{lie2form}}) follows from the identity
\begin{eqnarray}
\label{domega1}
d \omega = 3ig (X {\bf{e}}^5 - A) \wedge \omega
\end{eqnarray}
on using the gauge choice $g(-X+3A_5) =0$. Furthermore, we remark that the $pqr$ components of ({\ref{domega1}}) imply
({\ref{block3a1b}}), on contracting ({\ref{domega1}})
with $\bar{\xi}^p \xi^q \tau^r$.

Next we consider the components of $H$. On using ({\ref{hr5sol}}), $H_{r5}$ can be simplified to
\begin{eqnarray}
\label{hflux1}
H_{r5} &=& 2ig A_r -{1 \over 24} (\xi \wedge {\bar{\xi}})^{pq} 
\nabla_r \omega_{pq} + {1 \over 96} (d \omega)_{pn\ell}
\omega{}^p{}_m (\xi \wedge {\bar{\xi}})^{mn} (\xi \wedge {\bar{\xi}})_r{}^\ell
\nonumber \\
&-&{1 \over 12} \nabla^\mu \omega_{\mu q} (\xi \wedge {\bar{\xi}})_r{}^q
\end{eqnarray}

where we have made use of the identity
\begin{eqnarray}
{\bar{\tau}}^p \nabla_\ell \tau_p
&=&{1 \over 32}(d \omega)_{pn\ell}
\omega{}^p{}_m (\xi \wedge {\bar{\xi}})^{mn}
-{1 \over 4} \nabla^\mu \omega_{\mu \ell} \ .
\end{eqnarray}

In addition, the conditions ({\ref{block3d}}) and
({\ref{extrageocomp1}}) imply that

\begin{eqnarray}
\label{hflux2}
H_{pq} &=& i {\mathcal{F}} (\omega)_{pq}
-{1 \over 4} \big(\nabla^\mu \omega_{\mu 5}\big)
(\xi \wedge {\bar{\xi}})_{pq}
-{1 \over 3} (\xi \wedge {\bar{\xi}})_{[p}{}^\ell
\nabla_{|\ell|} \omega_{q]5}
\nonumber \\
&+& \big( -{i \over 3}gX -{1 \over 24} (\xi \wedge \bar{\xi})^{r \ell} \nabla_5 \omega_{r \ell} \big)
\omega_{[p}{}^\ell (\xi \wedge {\bar{\xi}})_{|\ell|q]}
\end{eqnarray}
where ${\cal{F}}$ is a real function which is not determined by the KSE.

\subsection{Gaugino KSE}

Next, we consider the gaugino KSE ({\ref{gaugino}}). These imply the following
conditions:

\begin{eqnarray}
\label{akse1}
{\cal{B}} (\epsilon, {\cal{A}}^I \epsilon)=0,
\end{eqnarray}
\begin{eqnarray}
\label{akse2}
{\cal{B}} (\epsilon, \gamma_\lambda {\cal{A}}^I \epsilon)=0,
\end{eqnarray}
\begin{eqnarray}
\label{akse3}
{\cal{B}} (\epsilon, \gamma_{\rho \sigma}{\cal{A}}^I \epsilon)=0  \ .
\end{eqnarray}
These conditions are both necessary and sufficient for ({\ref{gaugino}}) to hold,
the proof of sufficiency is given in Appendix C.

The condition ({\ref{akse1}}) is equivalent to
\begin{eqnarray}
\label{akse1a}
(F^I - X^I H)_{pq} \tau^p {\bar{\tau}}^q =0 \ .
\end{eqnarray}

The condition ({\ref{akse2}}) is equivalent
to

\begin{eqnarray}
i (\star \omega)_\lambda{}^{\mu \nu} (F^I - X^I H)_{\mu \nu} +2i \partial_\mu X^I (\omega)_\lambda{}^\mu =0 \ .
\end{eqnarray}

The case for which $\lambda=5$ holds automatically 
as a consequence of ({\ref{akse1a}}). The case for which
$\lambda =p$ implies that
\begin{eqnarray}
\label{akse2aa}
(\omega_1)_p{}^q \bigg( (F^I-X^I H)_{5q} + \partial_q X^I 
\bigg)=0
\end{eqnarray}
and hence
\begin{eqnarray}
\label{akse2a}
\tau^q \bigg((F^I -X^I H)_{5q} + \partial_q X^I \bigg)&=&0 \ .
\end{eqnarray}

Next, consider ({\ref{akse3}}), which is equivalent to
\begin{eqnarray}
\label{akse3intermed}
-2i (F^I -X^I H)_{\sigma \nu} \omega_{\rho}{}^\nu
+2i (F^I -X^I H)_{\rho \nu} \omega_{\sigma}{}^\nu
\nonumber \\
-2i (\star \omega)_{\rho \sigma}{}^\mu \partial_\mu X^I
+4g V_J (X^I X^J - {3 \over 2} Q^{IJ}) \omega_{\rho \sigma} =0 \ .
\end{eqnarray}

On setting $\sigma=5$, $\rho=p$ one obtains ({\ref{akse2aa}}). 
Finally, we set $\sigma=q$, $\rho=p$ in ({\ref{akse3intermed}}) we obtain

\begin{eqnarray}
\label{akse3re}
-2i (F^I -X^I H)_{q \ell} \omega_p{}^\ell
+2i (F^I -X^I H)_{p \ell} \omega_q{}^\ell
\nonumber \\
+ \bigg(4g V_J (X^I X^J - {3 \over 2} Q^{IJ}) 
+2i \partial_5 X^I \bigg) \omega_{pq} =0 \ .
\end{eqnarray}

On setting
\begin{eqnarray}
\Phi^I_q = -4i (F^I - X^I H)_{q \ell} {\bar{\tau}}^\ell
+ (4g V_J (X^I X^J - {3 \over 2} Q^{IJ})
+2i \partial_5 X^I) {\bar{\tau}}_q
\end{eqnarray}
the condition ({\ref{akse3re}}) is equivalent to
\begin{eqnarray}
\tau_p \Phi^I_q + {\bar{\tau}}_p {\bar{\Phi}}^I_q
- \tau_q \Phi^I_p - {\bar{\tau}}_q {\bar{\Phi}}^I_p =0 \ .
\end{eqnarray}
Furthermore $\tau^q \Phi^I_q = {\bar{\tau}}^q \Phi^I_q =0$ as 
a consequence of ({\ref{akse1a}}). It follows that
\begin{eqnarray}
\Phi^I_q = \Delta^I \tau_q + \Upsilon^I {\bar{\tau}}_q
\end{eqnarray}
where $\Upsilon^I - {\bar{\Upsilon}}^I =0$. Hence 
({\ref{akse3re}}) is equivalent to 

\begin{eqnarray}
\label{akse3re2}
-4i (F^I - X^I H)_{q \ell} {\bar{\tau}}^\ell
+ (4g V_J (X^I X^J - {3 \over 2} Q^{IJ})
+2i \partial_5 X^I) {\bar{\tau}}_q
\nonumber \\
= \Delta^I \tau_q + \Upsilon^I {\bar{\tau}}_q \ .
\end{eqnarray}
We remark that ({\ref{akse3re2}})
implies ({\ref{akse1a}}). Hence, ({\ref{akse1}}), ({\ref{akse2}}) and ({\ref{akse3}})
are equivalent to ({\ref{akse2a}}) and ({\ref{akse3re2}}).

The gaugino conditions are therefore equivalent to

\begin{eqnarray}
\label{gfincond}
F^I - X^I H &=& {{\bf{e}}}^5 \wedge (-d X^I + \alpha^I) + \Lambda^I
\nonumber \\
&+& \bigg(-{i \over 4} g V_J (X^I X^J - {3 \over 2} Q^{IJ}) + {1 \over 8} \partial_5 X^I \bigg)
\omega {\lrcorner}  (\xi \wedge {\bar{\xi}})
\end{eqnarray}
where  $\alpha^I$ are 1-forms satisfying
\begin{eqnarray}
\label{algg1}
\omega \wedge \alpha^I=0, \qquad X_I \alpha^I=0
\end{eqnarray}
and $\Lambda^I$ are 2-forms satisfying
\begin{eqnarray}
\label{algg2}
\omega \wedge \Lambda^I=0, \quad \big( \omega {\lrcorner}  (\xi \wedge {\bar{\xi}}) \big) \wedge \Lambda^I=0,
\qquad X_I \Lambda^I=0 \ .
\end{eqnarray}

\section{Examples}

In this section, we briefly consider some special classes of supersymmetric solutions.

\subsection{Solutions with $\nabla \omega =0$}

To begin, consider the special case when $\nabla \omega=0$. For such a case, the geometric
condition ${\widehat{\mathcal{N}}}=0$ holds automatically. Furthermore, the geometric condition
({\ref{lie2form}}) implies, on using $d \omega=0$, that $gX=0$. Thus, generically, solutions with parallel $\omega$ only exist for the ungauged theory, $g=0$. There is also considerable simplification to the $H$-flux conditions ({\ref{hflux1}}) and ({\ref{hflux2}}) which are equivalent to
\begin{eqnarray}
H = i {\mathcal{F}} \omega
\end{eqnarray}
where ${\mathcal{F}}$ is a real function. Furthermore the gaugino conditions ({\ref{gfincond}}) also simplify to give
\begin{eqnarray}
F^I = i X^I {\mathcal{F}} \omega + {{\bf{e}}}^5 \wedge (-d X^I + \alpha^I) + \Lambda^I
+{1 \over 8} \partial_5 X^I \omega {\lrcorner}  (\xi \wedge {\bar{\xi}})
\end{eqnarray}
where $\alpha^I$ and $\Lambda^I$ are 1- and 2-forms satisfying ({\ref{algg1}}) and ({\ref{algg2}})
respectively.

\subsection{Domain Wall Solutions}

We next consider consider domain wall solutions for which
all the gauge fields and potentials are set to zero. The geometric conditions
are equivalent to
\begin{eqnarray}
\nabla_\alpha \omega_{\mu \nu} = -igX (\star \omega)_{\alpha \mu \nu}
\end{eqnarray}
which is equivalent to requiring that $\omega$ be a Killing-Yano 2-form also
satisfying
\begin{eqnarray}
d \omega =-3igX \star \omega \ .
\end{eqnarray}
The conditions on the scalars are equivalent to 
\begin{eqnarray}
J (dX^I)^\sharp =0, \qquad {\cal{L}}_{e^5} X^I
= 2ig V_J (X^I X^J - {3 \over 2} Q^{IJ})
\end{eqnarray}
which can be rewritten as
\begin{eqnarray}
\omega \wedge dX^I+ 2ig V_J (X^I X^J - {3 \over 2} Q^{IJ}) \star \omega =0 \ .
\end{eqnarray}
In the ungauged theory, $\omega$ must be parallel, and the scalars satisfy $\omega \wedge dX^I =0$.

\subsection{A $N=6$ Supersymmetric Descendant Preon Solution}

Next, we shall explicitly construct a solution preserving exactly $N=6$ supersymmetry.
For the case of $N=6$ supersymmetry, we take three linearly independent (over $\mathbb{R}$) Majorana spinors $\{\epsilon_1, \epsilon_2, \epsilon_3\}$. These three spinors must be
orthogonal to a normal Majorana spinor $\psi$ with respect to ${\mathcal{B}}$. By using $Spin(3,2)$
gauge transformations, one may choose the normal spinor to be $\psi = 1+e_{12}$, and consequently
\begin{eqnarray}
\label{sixspan}
\epsilon_i \in {\rm span}_{\mathbb{R}} \{1+e_{12}, e_1 + e_2, i(e_1 - e_2) \} \ .
\end{eqnarray}

The solution we shall construct is an example of a descendant solution \cite{Gran:2007fu, Gran:2007kh}. In particular, for our solution, the gravitino equation preserves the maximal $N=8$ supersymmetry,
however the gaugino equation breaks the supersymmetry down from $N=8$ to $N=6$. 
For simplicity, we will work in the ungauged theory, setting $g=0$, and we also set the scalars
$X^I$ to be constant. However, we do not work in the minimal theory; in particular we take constants $Y^I$ such that $X_I Y^I =0$, and set $F^I = Y^I \chi$, for a real 2-form $\chi$, so that $H=0$. With this choice the gravitino equation simplifies to $\nabla \epsilon =0$, and we 
choose a trivial geometry consistent with this, which is $\mathbb{R}^{3,2}$, with
\begin{eqnarray}
ds^2 = (dx^1)^2 + (dx^2)^2 - (dx^3)^2 - (dx^4)^2 + (dx^5)^2 \ .
\end{eqnarray}
With this choice, it is clear that the gravitino equation admits maximal $N=8$ supersymmetry, and we take a Majorana basis of solutions to the gravitino equation consisting of the Majorana spinors $\epsilon_i$ for $i=1,2,3$ given in ({\ref{sixspan}}) together with $\epsilon_4 = i (1-e_{12})$.
Next we consider the gaugino equation corresponding to 
\begin{eqnarray}
\label{simpgaugino}
{\slashed{\chi}} \epsilon =0 \ .
\end{eqnarray}
Imposing ({\ref{simpgaugino}})  for $\epsilon = \epsilon_i$, $i=1,2,3$, we find that there exists a non-zero solution for $\chi$ corresponding to taking $\chi = iq \omega$ for $q \in \mathbb{R}$ constant, $q \neq 0$, and where $\omega$ is given by ({\ref{omdef2}}). So, it is clear that the gaugino equation admits a nontrivial 
flux $F^I \neq 0$ consistent with $N=6$ supersymmetry. Furthermore, the amount of (gaugino) supersymmetry
preserved by such a solution is exactly $N=6$, because imposing ({\ref{simpgaugino}})
for $\epsilon = \epsilon_i$, $i=1,2,3,4$ would force $F^I=0$. Furthermore, it is also
straightforward to verify that the all components of the Einstein and gauge field equations hold,
because in this case $\omega$ is parallel, and furthermore
\begin{eqnarray}
\omega \wedge \omega =0, \qquad \omega_{\mu \lambda} \omega_\nu{}^\lambda =0
\end{eqnarray}
and the scalar field equations follow as a consequence of the integrability condition
({\ref{scalint}}).

Hence this solution, with
$F^I = iq Y^I \omega$, preserves  $N=8$ supersymmetry from the perspective of the gravitino equation, but it is broken to exactly $N=6$ supersymmetry by the gaugino equation.

\section{Conclusion}

We have obtained the necessary and sufficient conditions for solutions of $(3,2)$ signature supergravity, with $ig \in \mathbb{R}$, coupled to arbitrary many vector multiplets, to preserve the minimal $N=2$ supersymmetry:

\begin{itemize}

\item[(i)] The geometric conditions are ${\widehat{{\mathcal{N}}}} =0$ where ${\widehat{{\mathcal{N}}}}$ is defined by 
({\ref{nijen1}}), as well as the condition
\begin{eqnarray}
d \omega = 3ig (X {\bf{e}}^5 - A) \wedge \omega \ .
\end{eqnarray}
We have shown that the condition ${\widehat{{\mathcal{N}}}} =0$ is equivalent to
\begin{eqnarray}
J[JX,Y] + J [X,JY] -[JX,JY] + \omega(X,Y) (\star d \star \omega)^\sharp =0
\end{eqnarray}
where $J$ is the nilpotent endomorphism $J: T {\mathcal{M}} \rightarrow T {\mathcal{M}}$, satisfying $J^2=0$, given by
\begin{eqnarray}
g(JX,Y)= -i \omega(X,Y) \ .
\end{eqnarray}
Moreover, $J$ is integrable, in the sense that if $X, Y \in {\rm Ker} J$ then $[X,Y] \in {\rm Ker} J$.
\item[(ii)] The components of the $H$-flux are given by ({\ref{hflux1}}) and ({\ref{hflux2}}).
In particular, we note that not the entirety of $H$ is fixed by the gravitino KSE as there is a term proportional to $\omega$ which is projected out by the KSE.
\item[(iii)] The conditions obtained from the gaugino KSE are ({\ref{gfincond}}), together with
({\ref{algg1}}) and ({\ref{algg2}}).
\end{itemize}

In terms of the integrability conditions of the KSE, it is clear from the integrability condition
({\ref{scalint}}), that if the gauge field equations hold, then supersymmetry implies that
the scalar field equations also hold. Furthermore, the integrability condition ({\ref{einstint}})
implies that if the gauge field equations hold, then we have
\begin{eqnarray}
\label{einstint2}
E_{\mu \nu} \gamma^\nu \epsilon =0
\end{eqnarray}
where $E=0$ is equivalent to the Einstein equations. As noted in Appendix $C$, the condition
({\ref{einstint}}) is equivalent to
\begin{eqnarray}
E_{\mu 5}=0, \qquad E_{pq} \omega^q{}_\ell=0
\end{eqnarray}
which is not sufficient to impose vanishing of all components of the Einstein equation. Specifically, all components of the Einstein equation are forced to vanish, with the exception of the 3 (real) components associated with $E_{\tau \tau}$ and $E_{\tau \bar{\tau}}$. We note that this is in contrast to the cases found for Lorentzian signature supergravity \cite{Gauntlett:2002nw}, where for the null spinor orbit only one real component of the 
Einstein equations was unfixed by KSE integrability conditions. Due to the preponderance of additional null directions in the $(3,2)$ signature theory, it is perhaps unsurprising that more components of the Einstein equation are unfixed when compared to the Lorentzian theory.

Having classified the $N=2$ solutions, it remains to examine the cases of enhanced supersymmetry 
for $N=4$, $N=6$ and $N=8$. In the case of maximal supersymmetry $N=8$, the 
gaugino equation ({\ref{gaugino}}) implies that
\begin{eqnarray}
\label{maxvan}
F^I = X^I H,  \qquad dX^I=0, \qquad g(X X_I -V_I)=0
\end{eqnarray}
and consequently such solutions reduce to solutions of the minimal supergravity.
It remains to then consider the integrability conditions of ({\ref{graveq}}) in 
the minimal theory, which can be read off from the integrability calculation in
\cite{Gauntlett:2003fk}. If $g \neq 0$ then maximal supersymmetry
requires that the coefficient multiplying
the identity matrix in this condition vanish, which sets all of the fluxes to zero. 
Such a solution is therefore maximally symmetric. In the case $g=0$ the analysis
of the gravitino integrability conditions is more complicated, and one might expect that
there be $(3,2)$ signature analogues of solutions such as $AdS_2 \times S^3$, as
well as more esoteric solutions such as the G\"odel solution found in
\cite{Gauntlett:2002nw}. 

For the case of $N=6$ supersymmetry, we have already constructed a simple descendant solution
which preserves exactly $N=6$ supersymmetry. This was possible because the conditions
imposed on the fluxes by the gaugino equation in $(3,2)$ signature turn out to be weaker
than the gaugino conditions found in the Lorentzian theory. Again, as mentioned previously in the context of integrability of the gravitino equation, the fact that such an algebraic condition
imposes weaker conditions for $N=6$ supersymmetry when compared to the case of standard signature
supergravities is unsurprising, due to the additional null directions present in the
$(3,2)$ signature. In the case of standard Lorentzian supergravity, it has been shown that all solutions preserving $N=6$ supersymmetry
are locally isometric to $N=8$ maximally supersymmetric solutions \cite{Grover:2006ps}, following \cite{Gran:2006cn}. However, it is possible to break this (global) $N=8$ supersymmetry by
taking appropriate quotients \cite{Figueroa-OFarrill:2007eny}. It would be interesting to fully
classify the $N=6$ solutions in $(3,2)$ signature. We remark in particular that the $N=6$ supersymmetric solution we constructed only exists in the non-minimal theory (e.g for the STU model). It remains to be determined if there exist $N=6$ solutions in the minimal theory.

Finally, for the case of $N=4$ supersymmetry, we take a pair of linearly independent (over $\mathbb{R}$) Majorana spinors $\{ \epsilon_1, \epsilon_2 \}$. By using $Spin(3,2)$ gauge transformations, we set $\epsilon_1 = 1 + e_{12}$. Then, by using ${\mathfrak{sl}}(2, \mathbb{R}) \ltimes {\mathfrak{n}}_3$ gauge transformations which leave $\epsilon_1$ invariant, one can write $\epsilon_2$ in one of two simple ${\mathfrak{sl}}(2, \mathbb{R}) \ltimes {\mathfrak{n}}_3$ gauge-inequivalent forms, as described in $\epsilon_2 = e_1 + e_2$ or $\epsilon_2 = i \alpha (1-e_{12})$ for $\alpha \in \mathbb{R}$. For the case with $\epsilon_2= e_1+e_2$, we can take a (non-Majorana) spinor given by
\begin{eqnarray}
\eta = \epsilon_1 + i \epsilon_2 = 1+e_{12} + i(e_1+e_2)
\end{eqnarray} 
and note that
\begin{eqnarray}
e^{{\pi \over 4}(\gamma_{12} + \gamma_{34})}
e^{-{\pi \over 4} \gamma_{52}} \eta = \sqrt{2} (1+e_2)
\end{eqnarray}
and so without loss of generality we can consider a (non-Majorana) spinor $\epsilon=1+e_2$.

For case with $\epsilon_2 = i \alpha(1-e_{12})$, we define non-Majorana spinors
\begin{eqnarray}
\eta_\pm = \epsilon_1 \pm i \epsilon_2 = (1 \mp \alpha).1 + (1 \pm \alpha) e_{12} \ .
\end{eqnarray}
Depending on the sign of $\alpha$, one of $\eta_\pm$ is $Spin(3,2)$ gauge equivalent
to the spinor $\eta = f.1$, for $f \in \mathbb{R}$. Hence for the $N=4$ solutions, we
can consider without loss of generality a (non-Majorana) spinor $\epsilon = f.1$ for $f \in \mathbb{R}$. We leave these more detailed considerations of the $N=4,6,8$ enhanced supersymmetry cases for future work.

\setcounter{section}{0}
\setcounter{subsection}{0}

\appendix{Clifford Algebra Conventions}

We begin with a split signature basis:
\begin{eqnarray}
{\bf{e}}^1, \quad {\bf{e}}^2, \quad {\bf{e}}^{\bar{1}}=({\bf{e}}^1)^*,
\quad {\bf{e}}^{\bar{2}}=({\bf{e}}^2)^*, \quad {\bf{e}}^5, \quad ({\bf{e}}^5)^* ={\bf{e}}^5
\end{eqnarray}
with respect to which the metric is
\begin{eqnarray}
\label{cframe}
ds^2 = 2 {\bf{e}}^1 {\bf{e}}^{\bar{1}}
-2 {\bf{e}}^2 {\bf{e}}^{\bar{2}} + ({\bf{e}}^5)^2 \ .
\end{eqnarray}
With respect to this basis we define
\begin{eqnarray}
\Gamma_1=\sqrt{2} i_{e_1}, \quad \Gamma_2 = \sqrt{2} i i_{e_2},
\quad \Gamma_{\bar{1}}= \sqrt{2} e_1 \wedge, 
\quad \Gamma_{\bar{2}}= \sqrt{2} i e_2 \wedge \ .
\end{eqnarray}
and 
\begin{eqnarray}
\gamma_5 \equiv  -\Gamma_{1 \bar{1} 2 \bar{2}} \ .
\end{eqnarray}

The gamma matrices act on
the space of Dirac spinors,
which consists of the complexified
span of $\{1, e_1, e_2, e_{12}=e_1 \wedge e_2\}$. We will find it useful to also work with a real spacetime basis
$\{ {\hat{{\bf{e}}}}^1, {\hat{{\bf{e}}}}^2, {\hat{{\bf{e}}}}^3, {\hat{{\bf{e}}}}^4, {\bf{e}}^5 \}$
with respect to which the metric is
\begin{eqnarray}
\label{reframe}
ds^2 = ({\hat{{\bf{e}}}}^1)^2+({\hat{{\bf{e}}}}^2)^2-({\hat{{\bf{e}}}}^3)^2-({\hat{{\bf{e}}}}^4)^2
+({\bf{e}}^5)^2
\end{eqnarray}
and take
\begin{eqnarray}
\gamma_1 = {1 \over \sqrt{2}}(\Gamma_1+\Gamma_{\bar{1}}),
\quad \gamma_2= {i \over \sqrt{2}}(\Gamma_1-\Gamma_{\bar{1}})
\end{eqnarray}
and
\begin{eqnarray}
\gamma_3= {1 \over \sqrt{2}}(\Gamma_2+\Gamma_{\bar{2}}),
\quad
\gamma_4= {i \over \sqrt{2}}(\Gamma_2-\Gamma_{\bar{2}}) \ .
\end{eqnarray}

We choose the orientation such that
\begin{eqnarray}
\label{orient}
\gamma_{\mu \nu \rho \sigma \tau} = \epsilon_{\mu \nu \rho \sigma \tau}, \quad
\gamma_{\nu \rho \sigma \tau} = \epsilon_{\nu \rho \sigma \tau \mu} \gamma^\mu, \quad
\gamma_{\rho \sigma \tau} = -{1 \over 2} \epsilon_{\rho \sigma \tau \mu \nu}
\gamma^{\mu \nu} \ .
\end{eqnarray}

With respect to the basis $\{1, e_{12}, e_1, e_2\}$, the gamma matrices
$\gamma_\mu$, $\mu=1,2,3,4,5$ act as
\begin{eqnarray}
\gamma_1 &=& \begin{pmatrix} & 0 \quad & {\mathbb{I}}
\cr & {\mathbb{I}} \quad & 0 \end{pmatrix},
\quad \gamma_2 = \begin{pmatrix} &0 \quad  \ \ &i \sigma_3
\cr &-i\sigma_3 \quad & 0 \end{pmatrix},
\nonumber \\
\gamma_3 &=& \begin{pmatrix} & 0 \quad & -\sigma_2
\cr & \sigma_2 \quad & 0 \end{pmatrix},
\quad 
\gamma_4 = \begin{pmatrix} & 0 \quad & -\sigma_1
\cr & \sigma_1 \quad & 0 \end{pmatrix},
\nonumber \\
\gamma_5 &=& \gamma_{1234} =  \begin{pmatrix} & {\mathbb{I}}  \quad & 0 \cr & 0 \quad & -{\mathbb{I}} \end{pmatrix}
\end{eqnarray}
with{\footnote{$\sigma_i \sigma_j = \delta_{ij} {\mathbb{I}} + i \epsilon_{ijk} \sigma_k$}}
\begin{eqnarray}
\sigma_1 = \begin{pmatrix} & 0 \quad & 1
\cr & 1 \quad & 0 \end{pmatrix}, \quad \sigma_2 = \begin{pmatrix} & 0 \quad & -i
\cr & i \quad & 0 \end{pmatrix}, \quad \sigma_3 = \begin{pmatrix} & 1 \quad & 0
\cr & 0 \quad & -1 \end{pmatrix} \ .
\end{eqnarray}

It is useful to note that
\begin{eqnarray}
\gamma_{12} = \begin{pmatrix} & -i \sigma_3  \quad & 0 \cr & 0 \quad & i \sigma_3 \end{pmatrix},
\qquad \gamma_{34} = \begin{pmatrix} & i \sigma_3 \quad & 0 \cr & 0 \quad & i \sigma_3 \end{pmatrix}
\end{eqnarray}
\begin{eqnarray}
\gamma_{13} = \begin{pmatrix} & \sigma_2  \quad & 0 \cr & 0 \quad & - \sigma_2 \end{pmatrix},
\qquad \gamma_{24} = \begin{pmatrix} & - \sigma_2 \quad & 0 \cr & 0 \quad & - \sigma_2 \end{pmatrix}
\end{eqnarray}
\begin{eqnarray}
\gamma_{14} = \begin{pmatrix} &  \sigma_1  \quad & 0 \cr & 0 \quad & - \sigma_1 \end{pmatrix},
\qquad \gamma_{23} = \begin{pmatrix} & \sigma_1 \quad & 0 \cr & 0 \quad & \sigma_1 \end{pmatrix} \ .
\end{eqnarray}

The charge conjugation matrix $C$ is given by
\begin{eqnarray}
C =  \begin{pmatrix} & \sigma_1 \quad & 0
\cr & 0 \quad & \sigma_1 \end{pmatrix}
\end{eqnarray}
and satisfies
\begin{eqnarray}
[C*, \gamma_\mu]=0
\end{eqnarray}
with
\begin{eqnarray}
C* 1=e_{12}, \quad C* e_{12}=1, \quad C*e_1 = e_2, \quad C* e_2 = e_1 \ .
\end{eqnarray}

A $Spin(3,2)$ invariant inner product ${\cal{B}}$ is given by
\begin{eqnarray}
\label{ginp}
{\cal{B}} (\epsilon, \eta)= \langle B \epsilon, \eta\rangle
\end{eqnarray}
where
\begin{eqnarray}
B = \begin{pmatrix} &  \sigma_3  \quad & 0 \cr & 0 \quad &  \sigma_3 \end{pmatrix}
\end{eqnarray}
satisfies
\begin{eqnarray}
B.1=1, \qquad B.e_{12}=-e_{12}, \qquad B.e_1 = e_1, \qquad B.e_2 = -e_2 \ .
\end{eqnarray}
This inner product satisfies
\begin{eqnarray}
{\cal{B}}(\epsilon, \gamma_\mu \eta) =  {\cal{B}}(\gamma_\mu \epsilon, \eta) \ , \quad
{\cal{B}}(\epsilon, \gamma_{\mu \nu} \eta) = - {\cal{B}}(\gamma_{\mu \nu} \epsilon, \eta) 
\end{eqnarray}
and also
\begin{eqnarray}
{\cal{B}}(C* \epsilon, \eta) = - {\cal{B}}(C* \eta, \epsilon) \ .
\end{eqnarray}
In particular, if both $\eta$ and $\epsilon$ are Majorana with $C* \epsilon = \epsilon$
and $C* \eta = \eta$ then ${\mathcal{B}}(\epsilon, \eta)$ is purely imaginary (or zero).
Furthermore, if $\chi$ is a $k$-form then
\begin{eqnarray}
{\slashed{\chi}}= \chi_{\mu_1 \dots \mu_k} \gamma^{\mu_1 \dots \mu_k} \ .
\end{eqnarray}
We also use the convention that if $\chi$, $\psi$ are 2-forms, then
\begin{eqnarray}
(\chi \lrcorner ~ \psi)_{\mu \nu} = \chi_{[\mu}{}^\lambda \psi_{|\lambda| \nu]} \ .
\end{eqnarray}

\appendix{Gravitino KSE Linear System}

In this Appendix, we list the linear system obtained from the gravitino equation ({\ref{graveq}})
(with $k=1$) acting on the Majorana spinor $1+e_{12}$.

\begin{eqnarray}
\label{glinsys}
2i \Omega_{1, {\bar{1}} {\bar{2}}}
+(-\Omega_{1, 1 {\bar{1}}} + \Omega_{1, 2 \bar{2}})
-{1 \over 2} H_{15} +{i} H_{\bar{2} 5} +{3i} g A_1 &=&0
\nonumber \\
 (\Omega_{1, 1 {\bar{1}}} - \Omega_{1, 2 \bar{2}}) -2i \Omega_{1,12}
-{3 \over 2} H_{15} +{3i} g A_1 &=&0
\nonumber \\
-\Omega_{1,51} -{i} \Omega_{1,5 \bar{2}}
-{3i \over 2} H_{1 \bar{2}} &=&0
\nonumber \\
-\Omega_{1,5 \bar{1}} +{i} \Omega_{1,52}
-H_{1 \bar{1}} -{1\over 2}H_{2 \bar{2}}
+{i\over 2} H_{12} -{i} gX &=&0
\nonumber \\
2i \Omega_{2, {\bar{1}} {\bar{2}}} + (-\Omega_{2,1 \bar{1}}
+ \Omega_{2, 2 \bar{2}}) +{i} H_{\bar{1} 5}-{1 \over 2} H_{25} +{3i} g A_2 &=&0
\nonumber \\
(\Omega_{2,1 \bar{1}} - \Omega_{2,2 \bar{2}}) - 2i \Omega_{2,12}
-{3 \over 2} H_{25}  +{3i} g A_2 &=&0
\nonumber \\
-\Omega_{2,51} -{i} \Omega_{2,5 \bar{2}}
-{i \over 2} H_{1 \bar{1}} -{i} H_{2 \bar{2}}
+{1 \over 2} H_{12} -{g} X &=&0
\nonumber \\
-\Omega_{2,5 \bar{1}} +{i} \Omega_{2,52} +
{3 \over 2} H_{\bar{1} 2} &=&0
\nonumber \\
2i \Omega_{5, {\bar{1}} {\bar{2}}} + (-\Omega_{5, 1 \bar{1}} + \Omega_{5, 2 \bar{2}})
+{i} H_{\bar{1} \bar{2}} +{1 \over 2} (-H_{1 \bar{1}} + H_{2 \bar{2}})
-{i} gX + {3i} g A_5 &=&0
\nonumber \\
-\Omega_{5,51} -{i} \Omega_{5,5 \bar{2}}
+H_{15} + {i} H_{\bar{2} 5} &=&0 \ .
\nonumber \\
\end{eqnarray}

\appendix{Gaugino KSE}

In this appendix, we prove that the conditions ({\ref{akse1}}), ({\ref{akse2}}) and
({\ref{akse3}}) imply ({\ref{gaugino}}). To see this, note that
as $\epsilon$ is Majorana, and $[C*, {\mathcal{A}}^I]=0$, it follows that
${\mathcal{A}}^I \epsilon$ is Majorana, and therefore
\begin{eqnarray}
{\mathcal{A}}^I \epsilon =  \lambda^I.1 + {\bar{\lambda}}^I e_{12}
+ \mu^I e_1 + {\bar{\mu}}^I e_2
\end{eqnarray}
where $\lambda^I$, $\mu^I$ are complex. We then observe that
\begin{eqnarray}
{\mathcal{B}} (\epsilon, {\mathcal{A}}^I \epsilon) &=& \lambda^I - {\bar{\lambda}}^I
\nonumber \\
{\mathcal{B}} (\epsilon, \gamma_1 {\mathcal{A}}^I \epsilon) &=& \mu^I - {\bar{\mu}}^I
\nonumber \\
{\mathcal{B}} (\epsilon, \gamma_3 {\mathcal{A}}^I \epsilon) &=& i(\mu^I + {\bar{\mu}}^I)
\end{eqnarray}
and it therefore follows that ({\ref{akse1}}), ({\ref{akse2}}) imply that 
$\mu^I=0$ and $\lambda^I = {\bar{\lambda}}^I$, and so
${\mathcal{A}}^I \epsilon = \lambda^I (1+e_{12}) = \lambda^I \epsilon$ for
$\lambda^I \in \mathbb{R}$.
Hence
\begin{eqnarray}
{\mathcal{B}} (\epsilon, \gamma_{\mu \nu} {\mathcal{A}}^I \epsilon) =
\lambda^I \omega_{\mu \nu}
\end{eqnarray}
and therefore ({\ref{akse3}}) also implies $\lambda^I=0$.
Hence, ({\ref{akse1}}), ({\ref{akse2}}) and
({\ref{akse3}}) imply ({\ref{gaugino}}).

By way of a corollary to this, note that if $V$ is a 1-form, then the condition
${\slashed{V}} \epsilon=0$ is equivalent to $i_V \omega=0$ and $i_V \star \omega=0$,
which in turn is equivalent to $V_5=0$ and $i_V \omega=0$. This observation is useful in considering the integrability conditions of the gravitino KSE.

\appendix{N=4 Canonical Majorana Orbits}

In this appendix we present the construction of canonical Majorana spinor orbits for 
solutions preserving $N=4$ supersymmetry. We take the first Majorana spinor to be
$\epsilon_1 = 1 + e_{12}$ and the second Majorana spinor to be
\begin{eqnarray}
\epsilon_2 = \lambda . 1 + {\bar{\lambda}} e_{12} + \mu e_1 + {\bar{\mu}} e_2
\end{eqnarray}
for $\lambda, \mu \in \mathbb{C}$. We apply ${\mathfrak{sl}}(2, \mathbb{R})
\ltimes {\mathfrak{n}}_3$ gauge transformations, which leave $\epsilon_1$ invariant,
to obtain the simplest possible form for $\epsilon_2$. In what follows, all gauge
transformations utilized are elements of ${\mathfrak{sl}}(2, \mathbb{R})
\ltimes {\mathfrak{n}}_3$.

To begin, we consider $\{ \gamma_{12}+\gamma_{34}, \gamma_{13}+\gamma_{24}, \gamma_{14}-\gamma_{23} \}$, which
leave $\{1, e_{12}\}$ invariant and act as $\{\sigma_1, \sigma_2, i \sigma_3\}$
on $\{e_1, e_2 \}$. Utilizing such gauge transformations we can without loss of generality take
either $\mu = 1$ or $\mu=0$, resulting in two possible cases
\begin{itemize}
\item[(i)] $\epsilon_2 = \lambda. 1 + {\bar{\lambda}} e_{12} + e_1 + e_2$
\item[(ii)] $\epsilon_2 = \lambda. 1 + {\bar{\lambda}} e_{12}$
\end{itemize}
Next, consider $-\gamma_{12} + \gamma_{34} + \gamma_{13} -\gamma_{24}$, which leaves 
$\{ e_1, e_2 \}$ invariant, and acts as
\begin{eqnarray}
e^{\theta (-\gamma_{12} + \gamma_{34} + \gamma_{13} -\gamma_{24})}
\lambda. 1 + {\bar{\lambda}} e_{12} = \lambda' .1 + {\bar{\lambda}}' e_{12}
\end{eqnarray}
where if $\lambda = a+ib$ for $a,b \in \mathbb{R}$, $\lambda' =a-4 \theta b + ib$.
Consequently, if $b \neq 0$ then such a gauge transformation can be used, without loss of generality to set $a=0$. We then find three cases
\begin{itemize}
\item[(a)] $\epsilon_2 = i \alpha (1-e_{12}) + e_1 + e_2$
\item[(b)] $\epsilon_2 = \alpha (1+e_{12}) + e_1 + e_2$
\item[(c)] $\epsilon_2 = i \alpha (1-e_{12})$
\end{itemize}
for $\alpha \in \mathbb{R}$, where we have discarded the possible case $\epsilon_2 = \alpha (1+e_{12})$ as the gravitino KSE implies $\alpha$ must be constant, and such a case reduces back to $N=2$ supersymmetry.

Next, we note that
\begin{eqnarray}
e^{\phi \gamma_5 (\gamma_1 - \gamma_4)} (\alpha (1+e_{12}) + e_1 + e_2) = (\alpha + 2 \phi) (1+e_{12}) + e_1 + e_2
\end{eqnarray}
and hence the elements in case $(b)$ are gauge equivalent to $e_1 + e_2$ which is a special subcase of (a). Hence we discard case $(b)$, and consider only cases $(a)$ and $(c)$.
We then observe that
\begin{eqnarray}
e^{\psi \gamma_5 (\gamma_2 - \gamma_3)} (i \alpha (1-e_{12}) + e_1 + e_2) 
= i \alpha (1-e_{12}) + (1-2 \alpha \psi) (e_1 + e_2) \ .
\end{eqnarray}
So, an $\epsilon_2$ in case $(a)$ with $\alpha \neq 0$ is gauge-equivalent to
a spinor in case $(c)$. Therefore, we have produced two possible simple canonical orbits 
corresponding to:
\begin{itemize}
\item[(A)] $\epsilon_2 = e_1 + e_2$
\item[(B)] $\epsilon_2 = i \alpha (1-e_{12})$, for $\alpha \in \mathbb{R}$
\end{itemize}
We finally remark that orbits $(A)$ and $(B)$ are ${\mathfrak{sl}}(2, \mathbb{R}) \ltimes {\mathfrak{n}}_3$ gauge inequivalent. This is because
${\mathcal{B}}(1+e_{12}, e_1 + e_2)=0$, whereas ${\mathcal{B}}(1+e_{12}, i \alpha (1-e_{12}))
= 2 i \alpha \neq 0$.

\section*{Acknowledgments}

DF was partially supported by the STFC DTP Grant ST/S505742. The work of W. A. Sabra is supported in part by the National Science Foundation under grant number PHY-1620505.

\section*{Data Management}

No additional research data beyond the data presented and cited in this work are needed to validate the research findings in this work.

\end{document}